\documentstyle[12pt]{article}
\begin{document}

\begin{titlepage}
\begin{center}

\bigskip

\bigskip

\vspace{3\baselineskip}

{\Large \bf  New exact cosmological solutions to Brans-Dicke gravity with a self-interacting scalar-field}

\bigskip

\bigskip

\bigskip

\bigskip

{\bf Olga Arias and Israel Quiros}\\
\smallskip

{ \small \it  
Physics Department. Las Villas Central University.\\
Santa Clara 54830. Villa Clara. Cuba}

\bigskip

{\tt  israel@mfc.uclv.edu.cu} 

\bigskip

\bigskip

\bigskip

\bigskip

\vspace*{.5cm}

{\bf Abstract}\\
\end{center}
\noindent

We derive a new parametric class of exact cosmological solutions to Brans-Dicke theory of gravity with a self-interacting scalar field and a barotropic perfect fluid of ordinary matter, by assuming a linear relationship between the Hubble expansion parameter and the time derivative of the scalar field. As a consequence only a class of exponential potentials and their combinations can be treated. The relevance of the solutions found for the description of the cosmic evolution are discussed in detail. We focus our discussion mainly on the possibility to have superquintessence behavior. 


\bigskip

\bigskip

\end{titlepage}

\section{Introduction}

Dark energy or missing energy is one of the contemporary issues the physics community is more interested in due, mainly, to a relatively recent (revolutionary) discovery that our present universe is in a stage of accelerated expansion\cite{pr}. This missing component of the material content of the universe is the responsible for the current stage of accelerated expansion and accounts for 2/3 of the total energy content of the universe, determining its destiny\cite{turner}. This is a new form of energy with repulsive gravity and possible implications for quantum theory and supersymmetry breaking. Among others, the new cosmology is characterized by the following features\cite{turner}: 1) Flat, critical density accelerating universe, 2) early period of inflation, 3) density inhomogeneities produced from quantum fluctuations during inflation, 4) composition of 2/3 dark energy, 1/3 dark matter and 1/200 bright stars, 5) matter content: $(29\pm 4)\%$ cold dark matter, $(4\pm 1)\%$ baryons and $\sim 0.3\%$ neutrinos. Besides, there is evidence (recent observation of SN1997ff at redshift $z=1.7$) that the present stage of accelerated expansion was preceded by an early period of decelerated expansion\cite{turner}. Any theoretical framework intended to describe our present universe should be compatible with these observational evidences.

A self-interacting, slowly varying scalar field, most often called quintessence, has been meant to account for the dark energy component. In the simplest models of quintessence, minimally coupled scalar fields are considered. However, extended quintessence models have also been considered\cite{faraoni,uzan,dftorres}. In these models the scalar field that accounts for the dark energy component is non-minimally coupled to gravity and, besides, is a part of the gravity sector. Brans-Dicke theory is a prototype for this kind. Motivated by the possibility to have superaccelerated expansion and, correspondingly, superquintessence solutions, this type of theories is now being considered (see, for instance, Ref.\cite{dftorres} and references therein). In particular, in Ref.\cite{dftorres}, the author considered different non-minimally coupled theories (among them the Brans-Dicke theory) and showed that, although superquintessence may arise in these models, its effect is vanishingly small, i. e., is far beyong the present observational possibilities. In that paper the author numerically integrated the field equations by using a computed code. Instead, it would be of interest to have a parametric class of exact solutions to treat this subject.

The aim of the present paper is, precisely, to derive a new parametric class of exact cosmological solutions to Brans-Dicke gravity theory with a self-interacting scalar field, by assuming a linear relationship between the Hubble expansion parameter and the time derivative of the scalar field. This type of relationship is a particular case of a more general one that has been used in Ref. \cite{chimento1} to find general solutions in $N+1$-dimensional theories of gravity with a self-interacting (minimally coupled) scalar field and also to derive general solutions when we have a self-interacting scalar field with exponential potential plus a free scalar field.\footnotemark\footnotetext{A method where a linear relationship between the field variables and/or their derivatives is assumed has also been used in \cite{acq} to derive 4d Poincare invariant solutions in thick brane contexts.} Also, it has been already used by us\cite{agq} to derive exact cosmological solutions to Einstein's gravity with two fluids: a barotropic perfect fluid of ordinary matter, together with a self-interacting scalar field fluid accounting for the dark energy in the universe.  The assumed relationship between the Hubble parameter and the time derivative of the scalar field is suggested by an implicit symmetry of the field equations. 

We point out that it is not neccessary to make any a priori assumptions about the functional form of the self-interaction potential or about the scale factor behavior. These are obtained as outputs of the assumed linear relationship between the Hubble expansion parameter and the time derivative of the scalar field, once one integrates the field equations explicitely. As a consequence only a class of exponential potentials and their combinations can be treated. However, this is not a serious drawback of the method. In effect, combinations of exponentials are interesting alternatives since these arise in more fundamental (particle) contexts: supergravity and superstring\cite{bcn}, where these types of potentials appear after dimensional reduction. The relevance of the solutions found for the description of the cosmic evolution will be discussed in some detail making emphasis in the possibility for superquintessence solutions to arise. This issue has been alredy studied numerically in Ref.\cite{dftorres}.  

We will be concerned with flat Friedmann-Robertson-Walker (FRW) cosmologies with the line element given by:

\begin{equation}
ds^2= -dt^2+a(t)^2 \delta_{ik} dx^i dx^k , 
\end{equation}
where the indexes $i,k = 1,2,3$ and $a(t)$ is the scale factor. We use the system of units in which $8\pi G=c=1$.


\section{The Method}

The Brans-Dicke field equations, in the presence of a self-interacting scalar field, are:

\begin{equation}
3H^2+3H\dot\varphi-\frac{\omega}{2}\dot\phi^2=e^{-\varphi}(\rho_m+V), 
\end{equation}

\begin{equation}
2\dot H+3H^2+\ddot\varphi+\dot\varphi^2+2H\dot\varphi+\frac{\omega}{2}\dot\varphi^2=e^{-\varphi}\{(1-\gamma)\}\rho_m+V, 
\end{equation} 

\begin{equation}
\ddot\varphi+\dot\varphi^2+3H\dot\varphi=\frac{e^{-\varphi}}{2\omega+3}\{(4-3\gamma)\rho_m+4V-2V'\}, 
\end{equation}
where we have introduced a new field variable $\varphi$ instead of the original $\phi=e^\varphi$. In these equations $\omega$ is the Brans-Dicke coupling parameter, $\gamma$ is the barotropic index of the fluid of ordinary matter, $V$ is the self-interaction potential and $H=\frac{\dot a}{a}$ is the Hubble expansion parameter. The dot accounts for derivative in respect to the cosmic time $t$ meanwhile the comma denotes derivative in respect to the scalar field $\varphi$. The energy density of the ordinary matter (cold dark matter plus baryons and/or radiation) is related with the scale factor through $\rho_m=\rho_{0,\gamma} a^{-3\gamma}$, where $\rho_{0,\gamma}$ is an integration constant coming from integrating the conservation equation.

Let us introduce a new time variable $dt=e^{\varphi/2}d\tau$. Then the field equations(2-4) can be rewritten in the following way:

\begin{equation}
3\bar H^2=\rho_m+\rho_\varphi, 
\end{equation}

\begin{equation}
2\dot{\bar H}+3\bar H^2=-P_m-P_\varphi, 
\end{equation} 

\begin{equation}
\ddot\varphi+\frac{1}{2}\dot\varphi^2+3\bar H\dot\varphi=\frac{4-3\gamma}{2\omega+3}\rho_m+\frac{2}{2\omega+3}(2V-V'), 
\end{equation}
where

\begin{eqnarray}
\rho_\varphi=\frac{\omega}{2}\dot\varphi^2-3\bar H \dot\varphi+V,\nonumber\\
P_\varphi=\ddot\varphi+\frac{\omega+1}{2}\dot\varphi^2+\bar H \dot\varphi-V,\nonumber\\
P_m=(\gamma-1)\rho_m.
\end{eqnarray}

Now the dot means derivative in respect to the new time variable $\tau$ and $\bar H=\frac{da/d\tau}{a}$. Combining of the above equations yields

\begin{equation}
\dot{\bar H}+\frac{1}{2}\bar H \dot\varphi+3\bar H^2=\frac{(2-\gamma)\omega+1}{2\omega+3}\rho_m+\frac{2\omega+1}{2\omega+3}V+\frac{1}{2\omega+3}V'.
\end{equation}

An implicit symmetry of the LHS of equations (7) and (9) is made explicit by the change $\bar H\rightarrow k\dot\varphi$, where $k$ is a constant parameter. Therefore, if we assume the linear relationship

\begin{equation}
\bar H=k\dot\varphi,\;\;\Rightarrow\; a=e^{k\varphi},
\end{equation}
then, the following differential equation for the self-interaction potential $V$ is obtained:

\begin{equation}
V'+\frac{2\omega+1-4k}{2k+1}V=\frac{(4-3\gamma)k-(2-\gamma)\omega-1}{2k+1}\rho_m.
\end{equation}

Straightforward integration of Eq. (11) yields

\begin{equation}
V(\varphi(a))=Aa^n+Ba^{-3\gamma},
\end{equation}
where

\begin{eqnarray}
n&=&\frac{4k-2\omega-1}{(2k+1)k},\nonumber\\
A=\frac{C_0}{3\gamma k-\alpha}&,&\;B=\frac{\beta\rho_{0,\gamma}}{3\gamma k-\alpha},\nonumber\\
\alpha=-kn,\;\;\beta&=&\frac{(4-3\gamma)k-(2-\gamma)\omega-1}{2k+1}.
\end{eqnarray}

An interesting feature of this potential is that it depends on the type of ordinary fluid which fills the universe. Otherwise, it depends on the barotropic index $\gamma$ of the matter fluid. This fact implies some kind of interaction between the ordinary matter and the quintessence field much like the interacting quintessence studied in Ref.\cite{dpavon}.

By substituting Eq. (12) back into the Friedmann equation (5) one obtains

\begin{equation}
\bar H^2=\bar A a^n+\bar B a^{-3\gamma},
\end{equation}
where

\begin{equation}
\bar A=\frac{2k^2 A}{6k^2+6k-\omega},\;\;\bar B=\frac{2k^2(\rho_{0,\gamma}+B)}{6k^2+6k-\omega}.
\end{equation}

Finally we make another change of time variable: $dr=a^{n/2} d\tau$ to get Eq. (14) integrated in quadratures:

\begin{equation}
\int\frac{a^{\frac{3\gamma+n}{2}-1}da}{\sqrt{\bar B+\bar A a^{3\gamma+n}}}=r+r_0,
\end{equation}
where $r_0$ is another integration constant.

\section{The Class of Solutions}

Straightforward integration of Eq. (16) leads to:

\begin{equation}
a(r)=a_0\{\sinh[\mu(r+r_0)]\}^\frac{2}{3\gamma+n},
\end{equation}
with

\begin{equation}
a_0=(\bar B/\bar A)^\frac{1}{3\gamma+n},\;\;\mu=\frac{3\gamma+n}{2}\sqrt{\bar A}.
\end{equation}

Then, from the linear relationship (10), one can derive the form of the scalar field:

\begin{equation}
\varphi(r)=\frac{1}{k}\ln a_0+\frac{2}{(3\gamma+n)k}\ln\sinh[\mu(r+r_0)].
\end{equation}

For purpose of observational testing we give the list of magnitudes of observational interest. The Hubble expansion parameter;

\begin{equation}
\bar H(r)=\frac{2\mu}{3\gamma+n}a(r)^{n/2}\coth[\mu(r+r_0)],
\end{equation}
the matter density parameter,

\begin{equation}
\Omega_m(r)=\frac{\rho_m}{3\bar H^2}=(\frac{3\gamma+n}{2\mu})^2\frac{\rho_{0,\gamma}}{3}\cosh^{-2}[\mu(r+r_0)],
\end{equation}
the scalar field density parameter,

\begin{equation}
\Omega_\varphi(r)=1-\Omega_m(r),
\end{equation}
the deceleration parameter,

\begin{equation}
q(r)=-1+\frac{1-nk}{2k}+\frac{3\gamma+n}{2}\Omega_m(r),
\end{equation}
and the standard (perhaps not precisely the physically meaningful\cite{dftorres}) equation of state of the scalar field,

\begin{equation}
\omega_\varphi=\frac{P_\varphi}{\rho_\varphi}=-1-n/3.
\end{equation}

This class of solutions depends (in principle) on six parameters: $\gamma$, $\omega$, $k$, $\rho_{0,\gamma}$, $C_0$ and $r_0$.

\section{Reducing the Space of Parameters}

A very crude analysis shows that $\Omega_\varphi$ should never be negative since, otherwise, this would imply coarse violations of the known energy conditions. Therefore $\Omega_m$ should never be greater than unity. Since $\Omega_m$ (Eq. (21)) is a maximum at $r=-r_0$ where

\begin{equation}
\Omega_m(-r_0)=(\frac{3\gamma+n}{2\mu})^2\frac{\rho_{0,\gamma}}{3},
\end{equation}
then by requiring that 

\begin{equation}
\Omega_m(-r_0)=1,\;\;\Rightarrow\;\frac{3\gamma+n}{2\mu}=\sqrt{\frac{3}{\rho_{0,\gamma}}}.
\end{equation}

If we apply the normalization in which $\rho_{0,\gamma}=1$, then $\frac{3\gamma+n}{2\mu}=\sqrt{3}$ and Eq. (26) implies:

\begin{equation}
C_0=\frac{(6\gamma k^2+(4+3\gamma)k-2\omega-1)(6k^2+6k-\omega)}{6k^2(2k+1)}.
\end{equation}

By considering Eq. (26) and evaluating Eq. (21) at present ($r=0$) one obtains:

\begin{equation}
r_0=\frac{1}{\mu}arccosh[\frac{1}{\sqrt{\Omega_m(0)}}],
\end{equation}
where, from observations, one may place $\Omega_m(0)=0.3$. 

After these considerations the space of parameters is reduced to the following free parameters: $\gamma$, $\omega$ and $k$. Besides the parameter $\gamma$ is fixed by the type of fluid dominating the given stage of the cosmic evolution. At present ($r=0$) $\gamma=1$, i. e., the universe is filled with a cold dust of dark matter plus baryons mainly.

We want to point out that all of the physical parameters of observational interest (equations (21-24)) can be given as functions of the redshift $z$ simply by replacing in the former equations

\begin{equation}
\Omega_m(r)\rightarrow\Omega_m(z)=\frac{(1+z)^{3\gamma+n}}{(1+z)^{3\gamma+n}+\sinh^2[\mu r_0]}.
\end{equation}

This form of writing will be useful for purpose of observational testing.

\section{Observational Testing}

The main observational facts we consider are the following\cite{turner}:

1.- At present ($r=0$) the expansion is accelerated ($q(0) < 0$).

2.- The accelerated expansion is a relatively recent phenomenon. Observations point to a decelerated phase of the cosmic evolution at redshift $z=1.7$. There is agreement in that transition from decelerated into accelerated expansion occurred at a $z \approx 0.5$\cite{triess}.

3.- The equation of state for the scalar field at present $\omega_\varphi\sim -1$ (it behaves like a cosmological constant).

4.- Although, at present, both the scalar (quintessence) field and the ordinary matter have similar contributions in the energy content of the universe ($\Omega_m(0)=0.3\;\Rightarrow\;\Omega_\varphi(0)=0.7$), in the past, the ordinary matter dominated the cosmic evolution,\footnotemark\footnotetext{The quintessence field $\varphi$ should be subdominant in the past as required by nucleosynthesis constraints\cite{fj}} meanwhile, in the future, the quintessence field will dominate (it already dominates) and will, consequently, determine the destiny of the cosmic evolution.

In fig.1 the evolution of both the matter and scalar field density parameters is shown for $\omega=1000$ and an arbitrary $k=530$ ($\gamma=1$), as function of the redshift $z$. We see that the evolution of both magnitudes meets the observational requirements o matter dominance (scalar field subdominance) at high redshift (in the past) and scalar field dominance at low redshift (at present and in the future). 

From fig.2, where the evolution of the deceleration parameter $q\;\&\;z$ is shown for the same $\omega=1000$ and $k=530$, we see that, at present ($z=0$), $q(0)=-0.55 < 0$ (accelerated expansion) meanwhile, in the past ($z > 0$) we had a stage of decelerated expansion. The transition from decelerated into accelerated expansion occurred at $z=0.67$, which is in acceptable agreement with the claimed value $z\approx 0.5$. Besides, the observational fact of decelerated expansion at redshift $z=1.7$ is fulfiled.

The equation of state is always a constant $\omega_\varphi=-1.00007$ for the former values of the free parameters $\omega$ and $k$, in good agreement with the accepted value. In general the solution found meets all of the above observational requirements (points 1-4) for a really wide range in the parameter space ($\omega,k$).

\section{Constrains from Superquintessence}

As pointed out in Ref.\cite{faraoni}, superquintessence behavior ($\omega_\varphi < -1$) is a feature of models of quintessence based on non-minimally coupled scalar-tensor theories including Brans-Dicke theory with a self-interacting scalar field. In Ref.\cite{dftorres}, however, it was shown that expected superacceleration effect is small enough as to escape the reach of present and future observations. We now proceed to analyze the question of superquintessence within the frame of the solution found.

We shall interested in $n \geq 0$ (Eq. (13)) in order to have superquintessence behavior (see Eq. (24)). This, in turn, implies $k > (2\omega+1)/4$. Since, from experiment $\omega \geq 500$\cite{will}, then $k > 250.25$ is a necessary condition to have superquintessence. Consider $n$ as function of $k$ ($\omega$ constrained by experiment):

\begin{equation}
n=\frac{4k-2\omega-1}{k(2k+1)},
\end{equation}
which is a maximum $n*$ at $k=(2\omega+1+\sqrt{(2\omega+1)(2\omega+3)})/4$, where

\begin{equation}
n*=\frac{2\sqrt{(2\omega+1)(2\omega+3)}}{\{\frac{2\omega+1}{2}(1+\sqrt{\frac{2\omega+3}{2\omega+1}})\}^2+(\frac{2\omega+1}{2})(1+\sqrt{\frac{2\omega+3}{2\omega+1}})}.
\end{equation}

For instance, for $\omega=500$, the maximum value for $n$ is $n*\approx 0.002$ (at $k\approx 500.75$) so, according to Eq. (24), the maximum expected superacceleration effect is of very small magnitude ($0.2\%$). For other values of $k > (2\omega+1)/4$, this effect would be even smaller. However, as discussed in details in Ref.\cite{dftorres}, the physically meaningful definition of the equation of state for the scalar field reads;

\begin{equation}
\tilde\omega_\varphi=-1-\frac{1}{3}\frac{\ln[\tilde\rho_\varphi(r)/\tilde\rho_\varphi(0)]}{\ln a(r)},
\end{equation}
instead of Eq. (24). In (32) the following redefinition for the scalar field energy density has been used:

\begin{equation}
\tilde\rho_\varphi=\rho_\varphi-3\bar H^2\{1-a_0^{-1/k}(\sinh[\mu(r+r_0)])^{-\frac{2}{(3\gamma+n)k}}\}.
\end{equation}

In fig.3 we show the behavior of the equation of state both, the usual (Eq. (24)) and the physically meaningful (Eq. (32))\cite{dftorres}, as function of the parameter $k$ for $\omega=1000$. Maximum expected superquintessence, in this case, is obtained for $k\sim 800$ (the standard equation of state) or $k\sim 1300$ (the physically meaningful equation of state). In both cases the effect is very small (of about 0.04$\%$).

\section{Conclusions}

We have found a new parametric class of exact cosmological solutions to Brans-Dicke theory of gravity with a self-interacting scalar field. A class of self-interaction potentials that is a combination of exponentials and that is dependent on the kind of matter fluid that fills the space-time arise. This potential suggests some kind of interaction between the quintessence field and the ordinary matter similar to the kind in Ref. \cite{dpavon}.

The flat universe solution found meets the observational constrains on the physical parameters, among them $\Omega_m$ ($\Omega_\varphi$), $\omega_\varphi$, $q$ and the experimental requirement that $\omega > 500$\cite{will}. It reflects the observational fact of a past decelerated, ordinary matter (including radiation) dominated universe and a future scalar field (dark energy) dominated universe. 

The analytic study of the class of solutions derived in the present paper confirm the results of Ref. \cite{dftorres}. Superquintessence behavior, although present for a given region in parameter space ($\omega,k$), is behind the reach of present and future observational possibilities.

\bigskip

{\bf Acknowledgments}

We thank our colleague Tame Gonzales for helpful comments and the MES of Cuba by financial support of this research.


\newpage

\begin{figure}[b]
\caption{The matter density parameter $\Omega_m$ (solid line) and the scalar field density parameter $\Omega_\phi$ (dotted line) as functions of the redshift $z$ for $\gamma=1$, $\omega=1000$ and an arbitrary $k=530$. We have used the normalization in which $\rho_{0,\gamma}=1$. We have considered the observational fact that, at present, $\Omega_m(0)=0.3$. It is seen an early stage when the contribution from the quintessence field was subdominant (high redshift). At present ($z=0$) both contributions from dust dark matter (plus baryons mainly) and from the scalar field are of the same order ($\Omega_m(0)=1/3$ while $\Omega_\varphi(0)=2/3$). In the future the quintessence field will be dominant.} 
\end{figure}

\begin{figure}[b]
\caption{The deceleration parameter $q$ is shown as function of the redshift $z$. At present $q_0=-0.55$. It is seen that there are both an early decelerated phase where $q > 0$ and a late accelerated expansion period ($q<0$). The transition from decelerated into accelerated expansion occurs at $z=0.67$.} 
\end{figure}

\begin{figure}[b]
\caption{The standard equation of state for the scalar field $\omega_\varphi$ and the physically meaningful scalar field state equation $\tilde\omega_\varphi$ are shown as functions of the free parameter $k$. We have chosen $\gamma=1$ and $\omega=1000$ for definiteness. It is seen that both curves attain a minimum at some $k$, i. e., a maximum superquintessence effect takes place at these $k$-values. This effect is very small and falls outside of the reach of present (and future) observational posibilities\cite{dftorres}.} 
\end{figure}

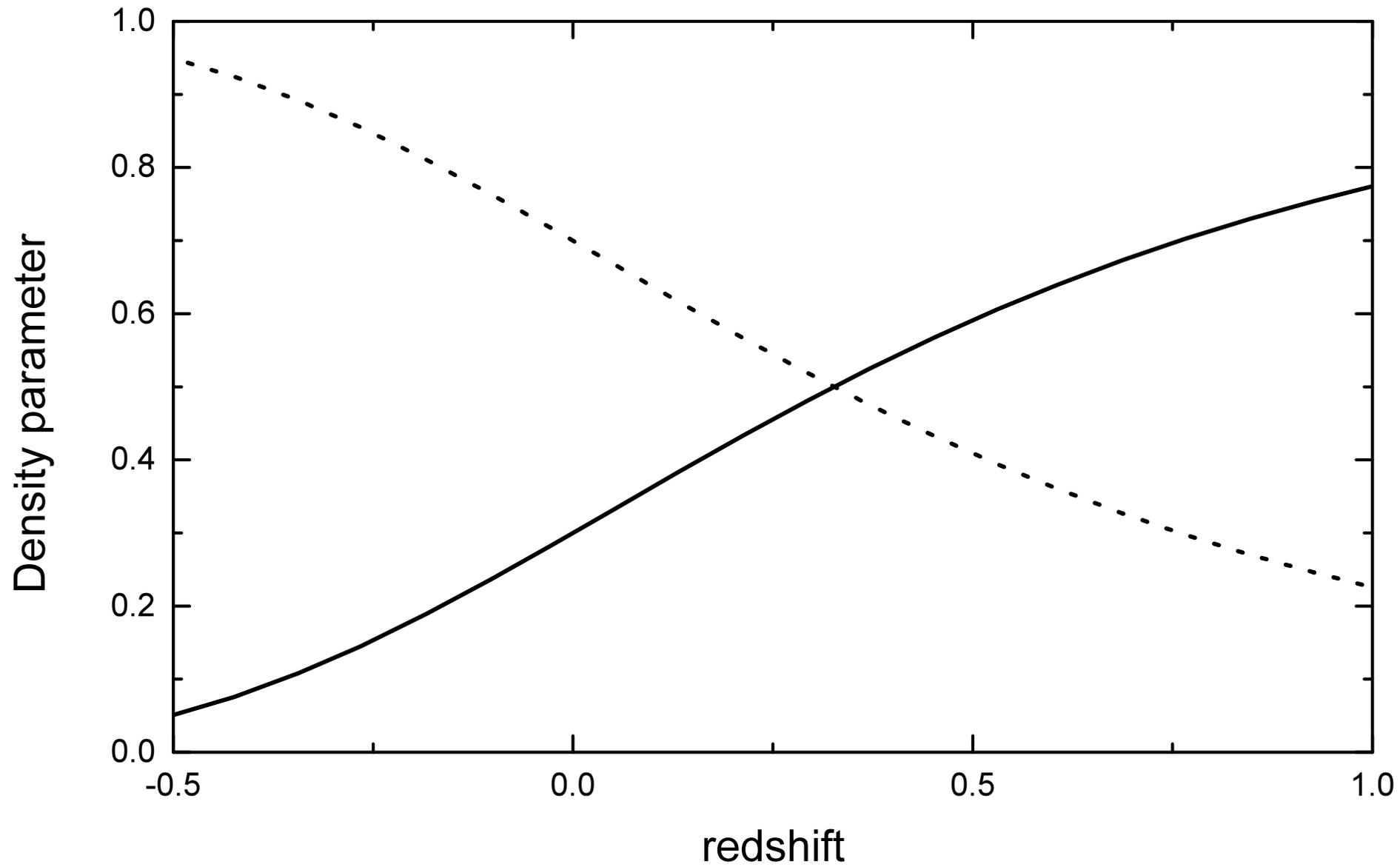
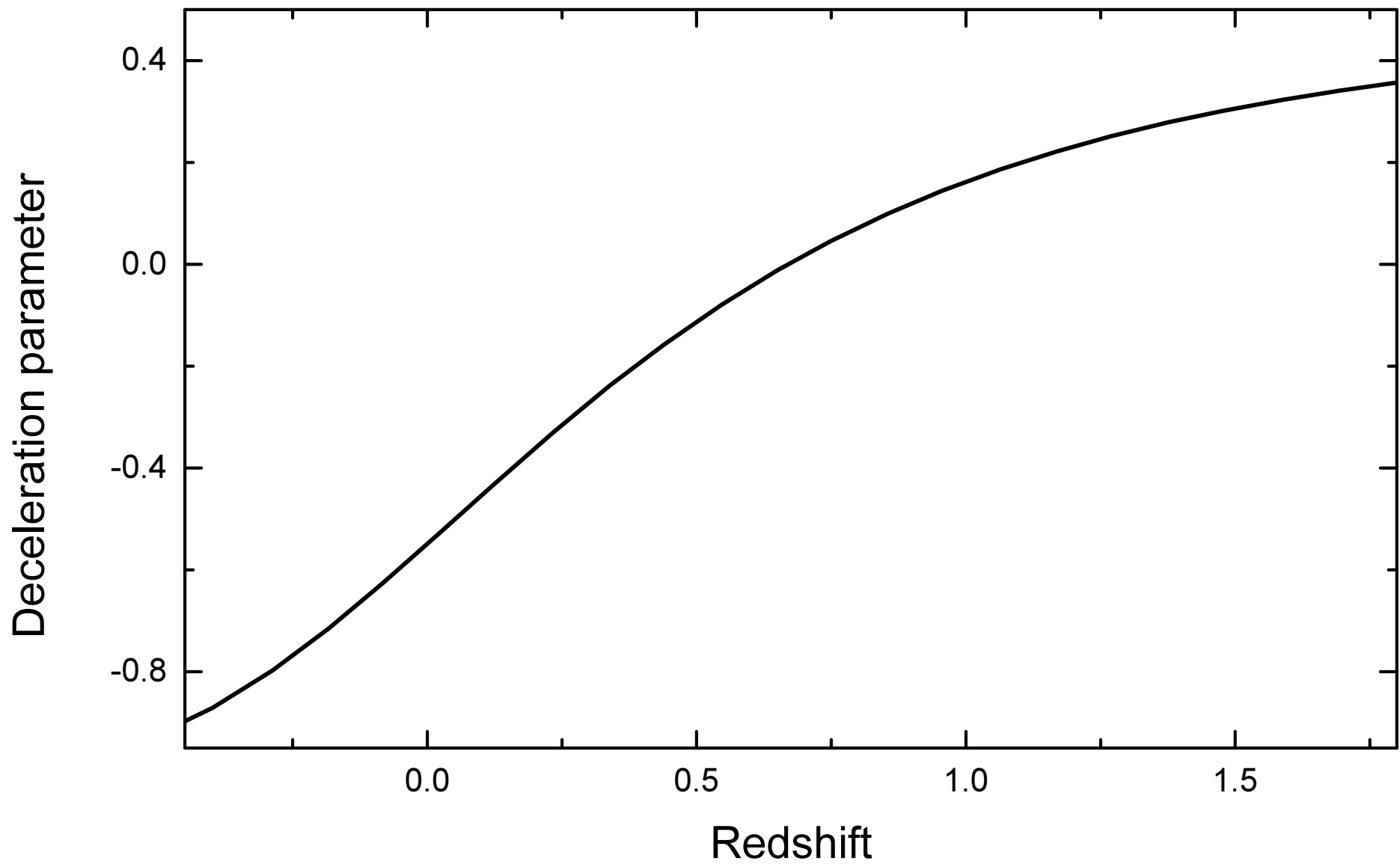
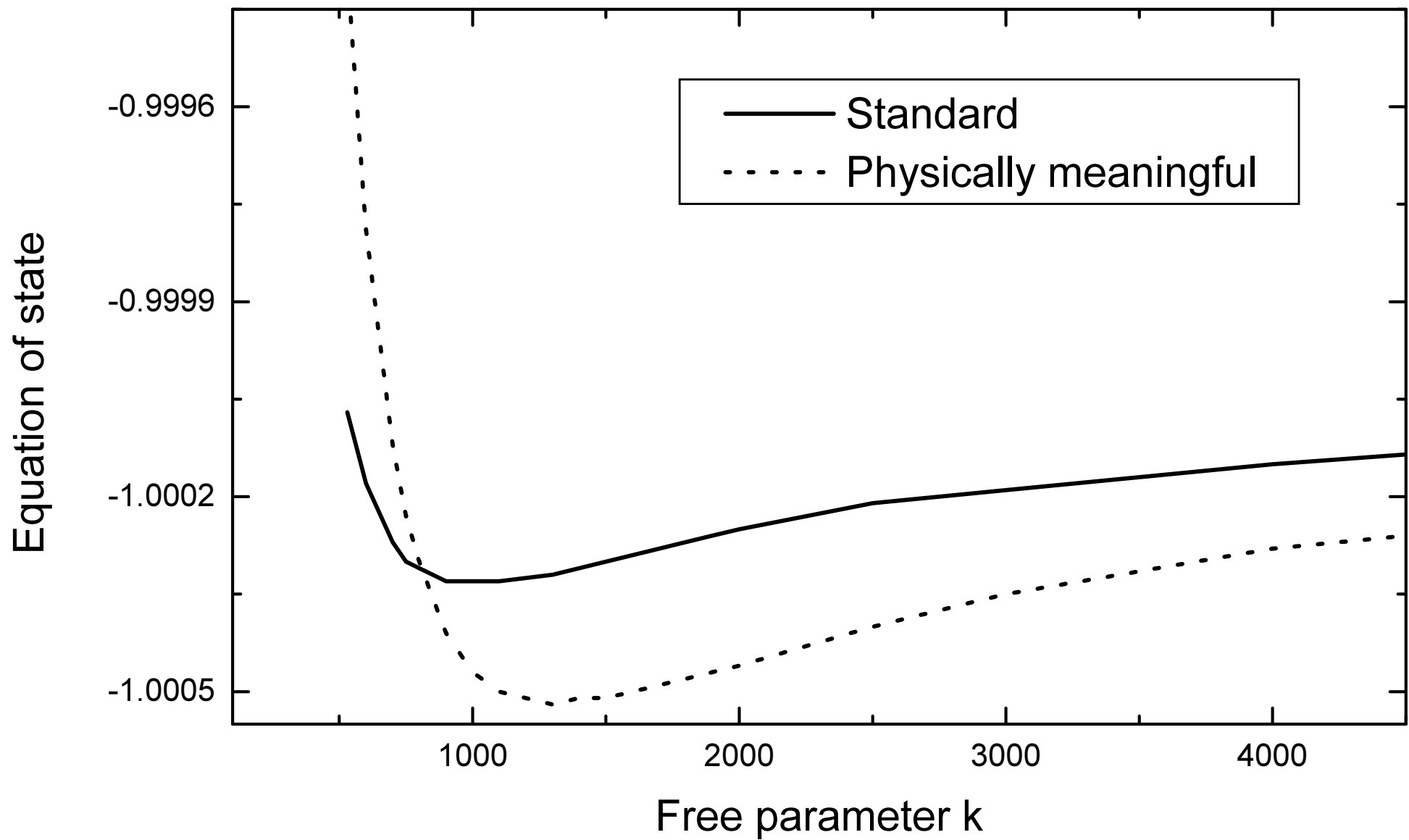
\end{document}